\documentclass[%
 reprint,
 amsmath,amssymb,
 aps, prl,
 longbibliography,
 twocolumn
]{revtex4-1}

\usepackage{graphicx}
\usepackage{dcolumn}
\usepackage{bm}
\usepackage{siunitx}
\usepackage{nicefrac}
\usepackage{mathrsfs}
\usepackage{dsfont}

\usepackage[plainpages=false,pdfpagelabels,colorlinks=true,linkcolor=blue,urlcolor=blue,citecolor=blue]{hyperref}

\begin{document}
	
\renewcommand{\vec}[1]{\mathbf{#1}}
\title{Topological pumping in a Floquet-Bloch band}

\author{Joaqu\'in Minguzzi}
\author{Zijie Zhu}
\author{Kilian Sandholzer}
\author{Anne-Sophie Walter}
\author{Konrad Viebahn}
\email{viebahnk@phys.ethz.ch}
\author{Tilman Esslinger}

\affiliation{Institute for Quantum Electronics, ETH Zurich, 8093 Zurich, Switzerland}

\date{\today}

\begin{abstract}
  Constructing new topological materials is of vital interest for the development of robust quantum applications. 
  However, engineering such materials often causes technological overhead, such as large magnetic fields, 
  spin-orbit coupling, 
  or dynamical superlattice potentials. 
  Simplifying the experimental requirements has been addressed on a conceptual level---by proposing to combine simple lattice structures 
  with Floquet engineering 
  ---but there has been no experimental implementation.
  Here, we demonstrate topological pumping in a Floquet-Bloch band using a plain sinusoidal lattice potential and two-tone driving with frequencies $\omega$ and $2\omega$. 
  We adiabatically prepare a near-insulating Floquet band of ultracold fermions via a frequency chirp, which avoids gap closings en route from trivial to topological bands.
  Subsequently, we induce topological pumping by slowly cycling the amplitude and the phase of the $2 \omega$ drive.
  Our system is well described by an effective Shockley model, 
  establishing a novel paradigm to engineer topological matter from simple underlying lattice geometries.
  This approach could enable the application of quantised pumping in metrology, 
  following recent experimental advances on two-frequency driving in real materials. 
\end{abstract}

\maketitle

The quantisation of charge transport in a Thouless pump shares its topological origin with the Quantum Hall effect, resulting from the Chern invariant defined on a torus geometry~\cite{thouless_quantization_1983}.
In case of the topological Thouless pump, the torus is spanned by one temporal and spatial dimension, requiring a one-dimensional insulator and a cyclic parameter change, respectively.
The topology of the insulator is determined by the Zak phase, which is the Berry phase for energy bands~\cite{zak_berrys_1989}.
An adiabatic change of the Zak phase by $2\pi$ induces charge transport of each particle by one unit cell.
Equivalently, this process can be understood as the shift of Wannier centres, i.e.~the electric polarisation~\cite{vanderbilt_berry_2018}, from one lattice site to the next.

\begin{figure*}[t]
	\begin{center}
		\includegraphics[width = 0.8\textwidth]{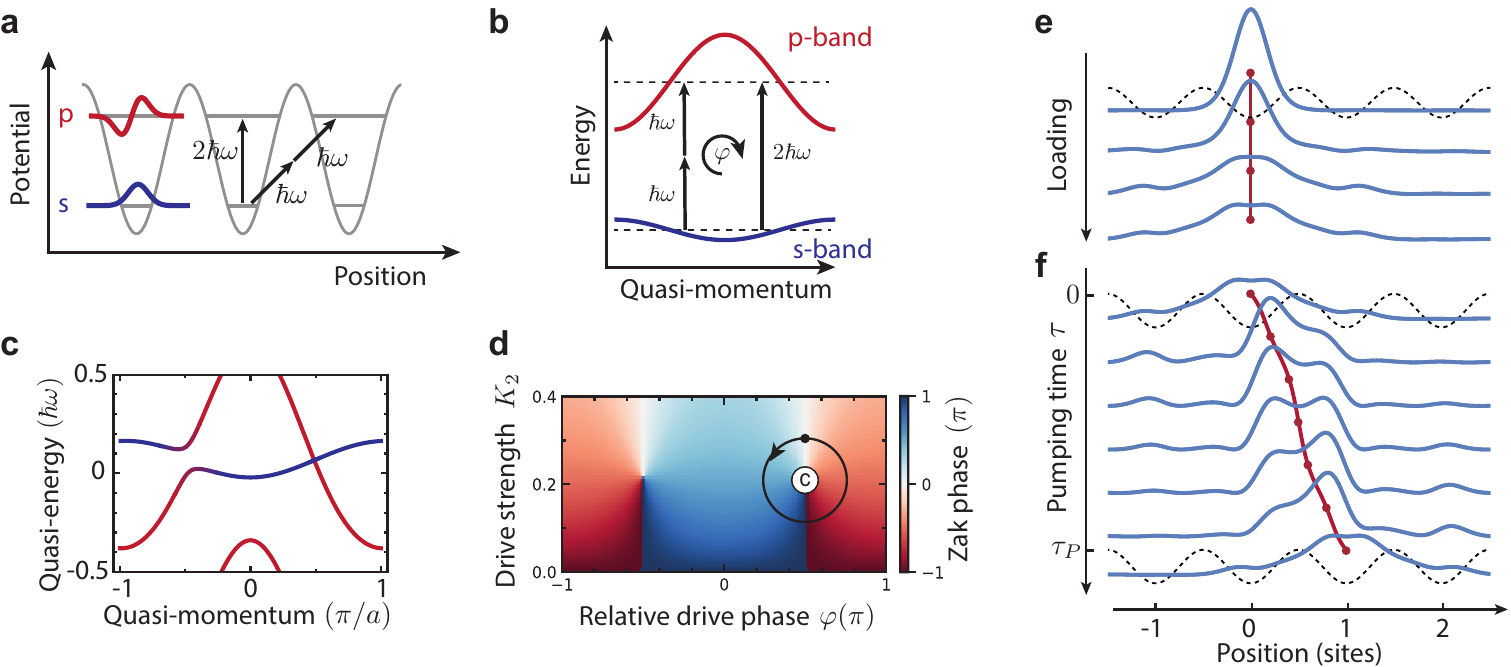}
		\caption{
			Floquet-Bloch bands and topological pump in a resonantly shaken optical lattice.
			\textbf{a} $s$- and $p$-orbitals of a sinusoidal potential are coupled on-site and between neighboring sites via one-photon ($2\omega$) or two-photon ($\omega$) resonances, respectively.
			\textbf{b} One- and two-photon resonances interfere constructively or destructively, depending on the relative phase $\varphi$ and quasi-momentum $q$.
			\textbf{c} Floquet-Bloch bands for critical parameters $\frac{\omega}{2\pi}=\SI{6.5}{kHz}$, $K_1=0.8$, $K_2=0.22$, $\varphi=+\pi/2$. The bands are asymmetric due to broken time-reversal symmetry and the band gap closes around quasi-momentum $q=+\frac{\pi}{2a}$ while it remains open at $q=-\frac{\pi}{2a}$.
			\textbf{d} Zak phase of the `lower' Floquet-Bloch band with topological transition points at $K_2=0.22$ and $\varphi=\pm\pi/2$. A counterclockwise orbit starting at $\varphi=+\pi/2$ (black dot) around the critical point marked as \textbf{c} (bandstructure in \textbf{c}) winds the Zak phase by $+2\pi$.
		    \textbf{e} During Floquet state preparation (vertical axis, `loading'), the initial $s$-band Wannier state (blue line, top) obtains on-site and neighbouring-site $p$-band contributions.
			\textbf{f} Topological pumping along the counterclockwise orbit shown in \textbf{d} (vertical axis, `Pumping time $\tau$') shifts the Wannier centre (red line) by one lattice site per pump period $\tau_P$.
		}
		\label{fig:1}
	\end{center}
\end{figure*}

So far, topological pumps have been realised in photonics~\cite{kraus_topological_2012,cerjan_thouless_2020,jurgensen_quantized_2021,ozawa_topological_2019} and ultracold atoms~\cite{lohse_thouless_2016,nakajima_topological_2016,lu_geometrical_2016,cooper_topological_2019}, while generalised pumps have also been studied using individual two-level systems~\cite{schroer_measuring_2014,ma_experimental_2018}.
Most implementations of topological pumps rely on dynamical superlattices to realise an effective Rice-Mele model~\cite{lohse_thouless_2016,nakajima_topological_2016,lu_geometrical_2016,cerjan_thouless_2020}.
However, the Rice-Mele pump is not suitable for real materials where the underlying lattice structure is essentially fixed.
In this work, we demonstrate a topological pump in a simple sinusoidal potential, enabled by Floquet engineering~\cite{oka_floquet_2019,rudner_band_2020,holthaus_floquet_2016,eckardt_colloquium:_2017,weitenberg_tailoring_2021}.
Resonant driving leads to an effective Shockley model~\cite{shockley_surface_1939,vanderbilt_berry_2018,velasco_classification_2019,fuchs_orbital_2021}, serving as a fundamentally different paradigm for topological pumping.
In contrast to the Rice-Mele model, which has two spatially separated lattice sites per unit cell, the two orbitals in the Shockley model are on the same lattice site.
This allows the unambiguous realisation of a topological transition in one dimension~\cite{fuchs_orbital_2021} and enables the appearance of fractionalised edge modes~\cite{velasco_realizing_2017}.

We use ultracold fermionic spin-polarised $^{40}$K atoms which are initially prepared in the $s$-band of a simple optical standing wave (Supplemental material).
The optical lattice is a sinusoidal potential $V(x)=V_0\cos^2(\pi x/a)$, with lattice spacing $a=\SI{532}{nm}$, depth $V_0=4\,E_r$ and recoil energy $E_r=h^2/8ma^2 =h\times \SI{4.41}{kHz}$, where $h$ is Planck's constant and $m$ is the mass of $^{40}$K.
The lattice is reflected off a piezo-actuated mirror, which is driven by an arbitrary waveform generator, resulting in a force $F(\tau)=\frac{\hbar\omega}{a}\left[K_1\cos(\omega\tau)+2K_2\cos(2\omega\tau+\varphi)\right]$.
This force on charge-neutral atoms is analogous to an effective $ac$ electric field for electrons in a solid.
The two-tone driving waveform~\cite{schiavoni_phase_2003,gommers_dissipation-induced_2005,zhuang_coherent_2013, niu_excitation_2015,grossert_experimental_2016,gorg_realization_2019,viebahn_suppressing_2021,yao_dynamics_2021} is parametrised by the dimensionless amplitudes $K_1$ and $K_2$ of the $\omega$ and $2\omega$ drives, respectively.
Each harmonic can hybridise the $s$ and $p$ bands (\hyperref[fig:1]{FIG. 1a-c})~\cite{viebahn_suppressing_2021,sandholzer_floquet_2022,zhang_shaping_2014,grossert_experimental_2016} leading to an effective two-band model with band inversion points (Shockley model~\cite{shockley_surface_1939,vanderbilt_berry_2018,velasco_classification_2019,fuchs_orbital_2021} or generalised Creutz ladder~\cite{sun_quantum_2017,kang_creutz_2020,kang_topological_2020}).
The relative phase $\varphi$ defines whether time-reversal symmetry is preserved ($\varphi = 0,\pi$) or broken ($\varphi\neq 0,\pi$).
In case of broken time-reversal symmetry the resulting Floquet-Bloch bands are asymmetric, enabling individual gap closings in the quasi-energy spectrum~\cite{sandholzer_floquet_2022}.
For $\varphi=\pm\pi/2$, the system fulfills a generalised space-time inversion symmetry~\cite{sandholzer_floquet_2022} and at a certain critical amplitude $K_2$ a single band gap closes. This is a topological transition point characterised by a Zak phase winding of $2\pi$ (\hyperref[fig:1]{FIG. 1d}), as pointed out by \citet{kang_topological_2020}.
If a single Floquet-Bloch band is uniformly filled, a topological pump will result from the cyclic variation of the driving parameters around the critical point (\hyperref[fig:1]{FIG. 1e,f}).
Contrary to previous works on directed group velocities due to time-reversal symmetry breaking~\cite{struck_tunable_2012,grossert_experimental_2016}, the Floquet-Thouless pump has a geometric origin~\cite{thouless_quantization_1983}.

\begin{figure}[t!]
	\begin{center}
		\includegraphics[width = 0.48\textwidth]{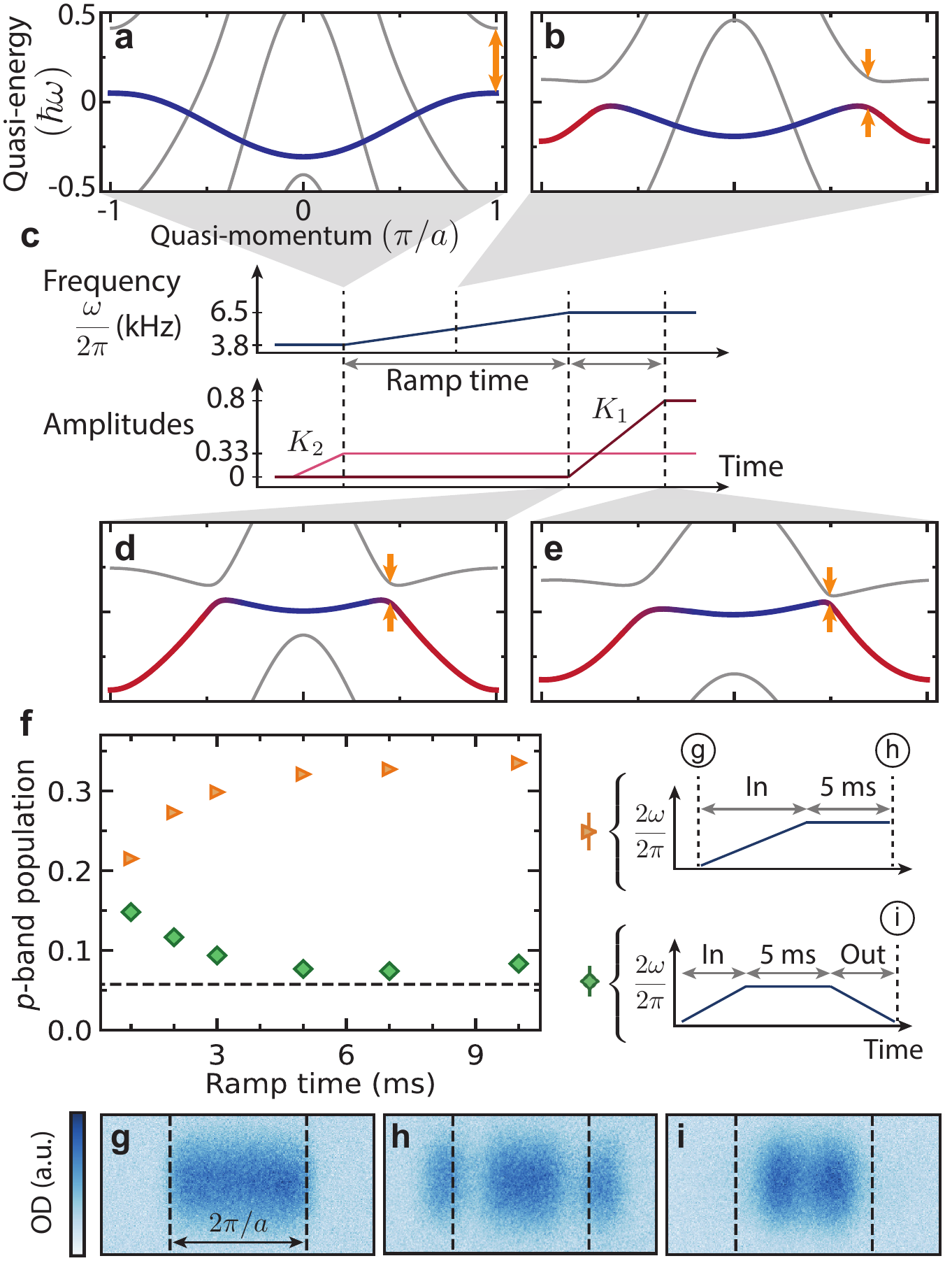}
		\caption{
			Adiabatic Floquet state preparation. \textbf{a-e} Floquet-Bloch quasi-energy bands and parameter ramps during the loading sequence. Occupied $s$ and $p$ bands are coloured in blue and red, respectively, while empty bands are displayed in grey. \textbf{a} Initially, the $s$-band (blue) is separated by a large gap from the $p$-band (orange arrow at the Brillouin zone (BZ) edge). \textbf{b},\textbf{d} During the frequency ramp (\textbf{c}) the minimal bandgap (orange arrows) `slides in' from the BZ edge and the outer portion of the Floquet band adiabatically changes from $s$- to $p$-type (red).
			\textbf{e} The target Floquet state is asymmetric in quasi-momentum at relative phase $\varphi = \pi/2$ due to broken time-reversal symmetry.
			During the loading sequence there is no gap closing.
			\textbf{f} Experimental frequency ramp optimisation, showing $p$-band population versus ramp time for $In$ (orange triangles) and $In$-$Out$ (green diamonds) sequences. Each data point is the mean and standard error of the mean (s.e.m.) of 3 repetitions; the error bar is smaller than the marker size. The horizontal dashed line indicates the static situation, in which $s$-band population is spuriously counted in the second BZ. \textbf{g-i} Raw optical density (OD) images after band mapping in the static lattice (\textbf{g}), after a \SI{5}{ms} $In$ sequence (\textbf{h}), and after a \SI{5}{ms} $In$-$Out$ sequence (\textbf{i}), respectively. The lattice potential is oriented horizontally, corresponding to the laboratory $x$ direction (Supplemental material). Vertical dashed lines mark the BZ edges.
		}
		\label{fig:2}
	\end{center}
\end{figure}

A key aspect of our work is the ability to connect the initially occupied $s$-band to the Floquet band without gap closing processes.
Here, we follow a three-step protocol that is outlined in \hyperref[fig:2]{FIG. 2}, motivated by ref.~\cite{kang_topological_2020}.
First, the $2\omega$ drive is switched on at a frequency red-detuned to the \emph{s-p} bandgap.
The dispersion of the occupied Floquet-Bloch band is similar to the dispersion of the static $s$-band.
Second, the $2\omega$ frequency is continuously increased, ending at its target value resonant to the \emph{s-p} bandgap.
During this stage the bands hybridise and the bandgaps move towards the BZ centre, reaching approximately $q=\pm\frac{\pi}{2a}$ at the end of the ramp.
Third, the amplitude $K_1$ is linearly increased from zero to its final value which creates an asymmetry between the two bandgaps if time-reversal symmetry is broken.
After the preparation, the minimal bandgap ($\SI{0.26}{kHz}\times h$) is located at approximately $q= +\frac{\pi}{2a}$ for $\varphi=+\pi/2$, as indicated by the orange arrows in \hyperref[fig:2]{FIG. 2}.

The topological pump relies on adiabatically connecting the static system to a single Floquet band.
The target band has $p$-band character at the BZ edges, shown in red in \hyperref[fig:2]{FIG. 2b,c,e}, whereas it has $s$-band character at the BZ centre, indicated in blue.
Thus, we can experimentally determine the adiabatic timescale for Floquet loading by counting $p$-band atoms in a band-mapping sequence and varying the frequency ramp duration followed by a hold time (Supplemental material).
The experimental data exhibits an increase of $p$-band population as function of ramp time, saturating at \SI{5}{ms} to a value of around $35\%$ (orange triangles in \hyperref[fig:2]{FIG. 2f}).
Exemplary band mapping images show $p$-band atoms in the outer parts of the BZ appearing at the expense of $s$-band atoms from the inner BZ (\hyperref[fig:2]{FIG. 2g,h}).
The measurements of $p$-band population suggest an adiabatic timescale of around \SI{5}{ms} which is consistent with an estimate from the Landau-Zener formula given by the minimal gap \SI{0.72}{kHz} to the `excited' Floquet band during the frequency ramp (Supplemental material).
We confirm this by reversing the loading process and transferring essentially all atoms back to the $s$-band for ramp times longer than \SI{5}{ms} (green diamonds, \emph{In-Out}).
While the loading protocol achieves adiabaticity with respect to band population, it also leads to redistribution of quasi-momentum states (\hyperref[fig:2]{FIG. 2i}).
The quasi-momentum dynamics is attributed to the expansion of the atomic cloud since we switch off the confining potential along the lattice direction prior to shaking (Supplemental material).
Similar measurements for the $K_1$ ramp produce an adiabatic timescale of 2 ms (Supplemental material).
In the following, we use ramp times of \SI{5}{ms} and \SI{2}{ms} for the $2\omega$ and $K_1$ ramps, respectively, during Floquet state preparation.
The optimised loading sequence initialises the Floquet state in the $(K_2,\varphi)$-plane at the points with $K_2=0.33$ and $\varphi=0,\pm\pi/2$, which are used as a starting point for all subsequent pumping orbits.

\begin{figure*}[t]
	\begin{center}
		\includegraphics[width = 0.66\textwidth]{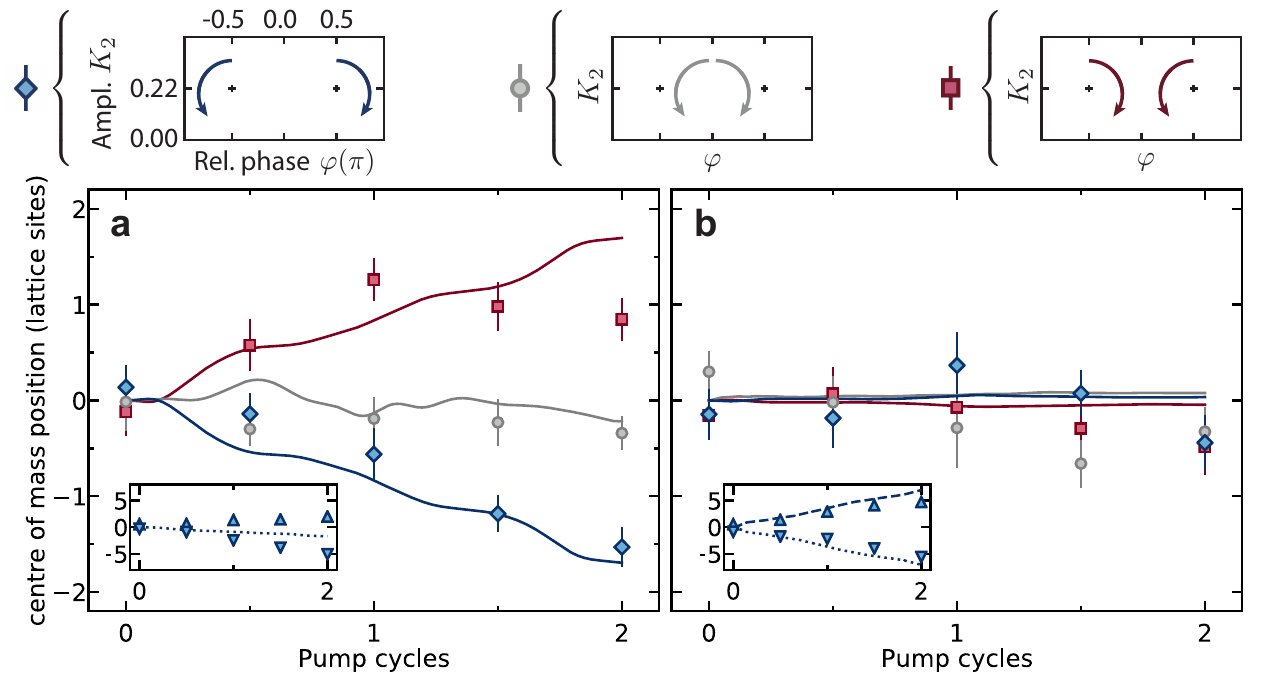}
		\caption{
			Topological pumping in a sinusoidal lattice potential. Detected \emph{in-situ} centre of mass (c.m.) position shown as function of pump cycles. Parameter cycles change the Zak phase by $-2\pi$ (blue), $0$ (grey) and $+2\pi$ (red). Each orbit is a sinusoidal modulation of $K_2$ and $\varphi$ centered around $K_2=0.20$ and $\varphi=0,\pm\pi/2$, with amplitudes $0.13$ and $\pi/4$, respectively. The other driving parameters are $\frac{\omega}{2\pi}=\SI{6.5}{kHz}$ and $K_1=0.8$. \textbf{a}-\textbf{b} Averaged c.m.~shifts for paired orbits with a pump period of \SI{4}{ms} (blue diamonds, grey circles and red squares). The initial state in \textbf{a} is the Floquet state prepared with the loading sequence (\hyperref[fig:2]{FIG. 2}) while in \textbf{b} the pumping waveform is suddenly switched on. Insets show c.m.~shifts for clockwise (blue inverted triangles and dotted lines) and counterclockwise (blue upright triangles and dashed lines) blue orbits. The dashed and dotted lines overlap in the inset of panel \textbf{a}. Each data point of a pumping orbit is the mean and s.e.m.~of 12 repetitions. Solid, dashed and dotted lines are numerical simulations. Theory and experiment differ from quantised pumping due to non-adiabatic effects, while in the experiment deviations from homogeneous band filling can additionally affect the pumping efficiency.
		}
		\label{fig:3}
	\end{center}
\end{figure*}

We demonstrate pumping by cycling the drive parameters around topological transition points in the $(K_2,\varphi)$-plane and recording the \emph{in-situ} centre-of-mass (c.m.) position of the atomic cloud (Supplemental material).
The lattice loading and the state preparation sequence can lead to a deviation from a homogeneous distribution of quasi-momentum states in the Floquet-Bloch band (\hyperref[fig:2]{FIG. 2}i).
Thus, group velocity effects can cause (non-topological) directional  motion due to the asymmetry in the band structure, similar to refs.~\cite{struck_tunable_2012,grossert_experimental_2016}.
To overcome this, the c.m.~position shift is averaged over paired orbits designed to cancel group velocity effects while inducing the same Zak phase winding, as illustrated in \hyperref[fig:3]{FIG. 3}.
For example, the blue clockwise and counterclockwise orbits around $\varphi=+\pi/2$ and $-\pi/2$, respectively, each change the Zak phase by $-2\pi$ but exhibit opposite band asymmetry.
In the experiment, we measure an average c.m.~shift for this pair of orbits consistent with topological pumping (blue diamonds in \hyperref[fig:3]{FIG. 3a}).
For the red orbits, which wind the Zak phase by $+2\pi$, we observe the opposite c.m.~shift (red squares) whereas the grey orbits, which give no net Zak phase winding, lead to no displacement (grey circles).
These results are in agreement with numerical simulations (blue, grey and red solid lines), based on solving the time-dependent Schr{\"o}dinger equation with an effective two-band Hamiltonian, assuming a homogeneous distribution of quasi-momentum states (Supplemental material).
Non-adiabatic effects and deviations from a homogeneous population of the band result in a below unit efficiency.
In contrast to previous realisations of topological pumps in superlattices~\cite{lohse_thouless_2016,nakajima_topological_2016}, the filled band remains dispersive throughout the pumping process.
Therefore, individual orbits can lead to significant motion of the atoms (approximately 5 lattice sites, \hyperref[fig:3]{FIG. 3a inset}), as expected for an imperfect band insulator with high tunnelling rates.

\begin{figure}[h!]
	\begin{center}
		\includegraphics[width = 0.5\textwidth]{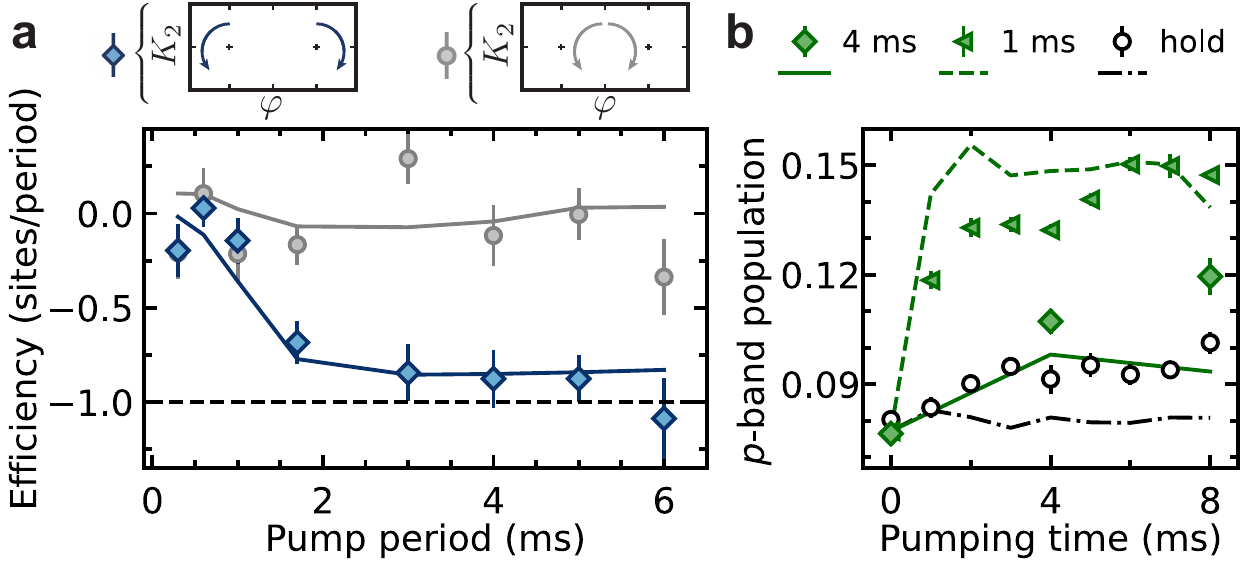}
		\caption{
			Adiabatic timescale for topological pumping. \textbf{a} Efficiency versus pump period for trivial (grey circles) and topological (blue diamonds) trajectories. Each data point is the mean and error of the fitted slope to a c.m.~shift versus pump cycles (datasets such as \hyperref[fig:3]{FIG.~3a}). Blue and grey solid lines are numerical simulations. These predict quantised efficiency of $-1.0$ for periods longer than \SI{12}{ms} as shown in Fig.~S1  in the supplemental material. \textbf{b} $In$-$Out$ measurement of $p$-band population versus pumping time for a topological orbit with pump periods \SI{4}{ms} (green diamonds), \SI{1}{ms} (green triangles) and hold (black circles). Each data point is the mean and s.e.m. of 3 repetitions. Solid, dashed and dashed-dotted lines are numerical simulations for the \SI{4}{ms}, \SI{1}{ms}, and hold, respectively.
		}
		\label{fig:4}
	\end{center}
\end{figure}

To highlight the importance of the loading protocol, we complement our previous measurements with identical parameter cycles but without adiabatic state preparation, i.e.~with a sudden switch-on of the drives at the beginning of the each orbit (\hyperref[fig:3]{FIG. 3b}).
In such an experiment we observe no average c.m.~shift for any pair of orbits (blue, red, or grey).
Numerical simulations assuming an initial $s$-band insulator agree not only with the averaged shifts (blue, grey and red solid lines) but also with the individual shifts induced by the clockwise and counterclockwise orbits (\hyperref[fig:3]{FIG. 3b inset}).

Adiabatic state evolution is key to sustain particle transport in a topological pump.
We investigate the pump adiabaticity by measuring its efficiency for different pump periods, shown in \hyperref[fig:4]{FIG. 4a}.
For each value of pump period, we extract the efficiency, defined as the shift in lattice sites per period, by fitting lines to datasets such as \hyperref[fig:3]{FIG. 3}.
Trivial orbits yield efficiencies around zero for all measured pump periods (grey circles).
In contrast, the efficiency for topological orbits increases from zero for periods below \SI{1}{ms} to almost unity for periods above \SI{3}{ms}, saturating at around -0.85 sites per period (blue diamonds).
This timescale for adiabatic transport is consistent with the inverse of the minimal gap size $(\SI{0.26}{kHz})^{-1}\approx\SI{3.9}{ms}$, representing an estimate for diabatic transfers to the `excited' Floquet band induced by the pump~\cite{kang_topological_2020}.
Numerical simulations (blue and grey lines) corroborate near-unit efficiency after 3 ms, while quantised efficiency is predicted for periods longer than \SI{12}{ms} (Fig.~S1).
Residual magnetic and optical potential gradients, inhomogeneous filling, as well as $d$-band excitations constitute the main limitations for reaching the fully quantised, adiabatic limit (Supplemental material).

We perform additional band population measurements to verify the adiabatic timescale for topological transport (\hyperref[fig:4]{FIG. 4b}).
Specifically, we employ an \emph{In-Out} sequence, similar to the one shown in \hyperref[fig:2]{FIG. 2f}, to detect diabatic transfers due to the pump as accumulated $p$-band population.
Here, the shaking waveform includes loading, pumping, and a time-reversed loading protocol that connects back to the static lattice.
We compare the accumulated $p$-band population between three different cases: pumping with periods of \SI{1}{ms} (green triangles) and \SI{4}{ms} (green diamonds), as well as shaking with fixed amplitude $K_2$ and relative phase $\varphi$ (black circles, `hold').
The $p$-band population builds up faster for the \SI{1}{ms} case than for the \SI{4}{ms} period, in accordance with measured transport efficiencies of -0.14(12) and -0.88(15) sites per period, respectively.
In turn, the \SI{4}{ms} period shows larger $p$-band population compared to the hold case, which indicates that a fully adiabatic pump is expected for longer periods, in agreement with numerical simulations (lines).
The small rise of $p$-band population measured in the no pump (hold) case, which is not predicted by simulations, is attributed to the expansion dynamics of the atomic cloud.
Discrepancies in the absolute value of $p$-band population between theory and experiment can result from an inhomogeneous distribution of quasi-momentum states (Supplemental material).
Overall, the band population measurements support the \emph{in-situ} observation of topological transport in the Floquet-engineered Thouless pump.

In conclusion, we demonstrated adiabatic Floquet state preparation and topological pumping of ultracold fermions in a simple sinusoidal potential.
The employed loading strategy improves existing methods to prepare topological Floquet states, as the topological gap `slides in' from the BZ edge instead of opening directly at the band inversion~\cite{dauphin_loading_2017}.
Moreover, the conceptual simplicity of the driving scheme presented here should be applicable to two-dimensional lattices~\cite{zhang_shaping_2014,zheng_floquet_2014,baur_dynamic_2014,phong_optically_2019}, establishing a new direction for topological Floquet engineering.
Finally, we demonstrated a topological pump which does not rely on superlattice potentials, possibly enabling the metrological application of quantised currents~\cite{pekola_single-electron_2013} in real materials via two-tone driving~\cite{heide_optical_2021}.

\appendix
\section{Acknowledgments}
We thank Nigel Cooper, Davide Dreon, Marius Gächter, Martin Holthaus, Gregor Jotzu, Daniel Malz, Stephan Roschinski, and Oded Zilberberg for inspiring discussions and comments on the manuscript.
We would like to thank Alexander Frank for his contributions to the electronic setup of the experiment.
K.V.~is supported by the ETH Fellowship programme.
This work was partly funded by the SNF (project nos. 169320 and 182650), NCCR-QSIT, QUIC (Swiss State Secretary for Education, Research and Innovation contract no. 15.0019), and ERC advanced grant TransQ (project no. 742579).


\section{Author Information} 
The authors declare no competing financial interests. Correspondence and requests for materials should be addressed to K.V.

%

\newpage
\cleardoublepage

\setcounter{figure}{0} 
\setcounter{equation}{0} 

\renewcommand\theequation{S\arabic{equation}} 
\renewcommand\thefigure{S\arabic{figure}} 
\setcounter{page}{1}

\part{Supplemental material}

\section{\large I. Experimental techniques}
\label{sec:suppI}
\subsection{General experimental sequence}
The experiment starts with evaporately cooled $350(50)\times10^3$ ultracold $^{40}$K fermionic atoms spin-polarised in the magnetic state $F=9/2,m_F=-9/2$. The magnetisation axis is defined by a low homogeneous magnetic field of \SI{10}{G} along the spatial $x$ direction. The atoms are initially confined in a crossed dipole trap along the $x$ and $y$ directions (gravity points towards the $-z$ direction) using \SI{826}{nm} laser light. This configuration provides confinement along the $x$ direction with an estimated harmonic trapping frequency of \SI{30}{Hz}. To achieve a near-insulating $s$-band state in the static lattice, the optical lattice along $x$ direction is switched on in \SI{200}{ms} with an s-shape ramp up to a lattice depth of $6\ E_R$. At the same time, we use s-shape ramps to switch-off the dipole trap along $y$ direction and ramp up the power of the dipole trap along $x$ direction. Subsequently, the lattice power is ramped down linearly from $6$ to $4\ E_R$ in \SI{10}{ms}. The final configuration provides confinement along $y$ and $z$ directions with harmonic trapping frequencies of \SI{64}{Hz} and \SI{157}{Hz}, respectively. We measured an upper bound of \SI{4}{Hz} for the trapping frequency along the $x$ direction. This preparation gives a better band insulating state than directly loading a $4\ E_R$ lattice, together with the weakest possible confinement along the $x$ direction. Both properties were found to be relevant to achieve larger $p$-band loading and also near-quantised pumping, see \hyperref[sec:suppIII]{III. Additional measurements}.

The optical lattice is formed by a single retro-reflected  $\lambda=\SI{1064}{nm}$ laser beam. We implement lattice shaking by modulating the position of the retro-reflecting mirror periodically in time (shaking waveform) using a piezo-electric actuator. The switch-on/off of a shaking waveform consists in a linear ramp of the amplitude in \SI{0.3}{ms}. This amounts to about 1.75 cycles for the slowest frequency used in the experiment. For example, it is used at the initial ramp of $K_2$ from $0$ to $0.33$ at a frequency $2\omega/2\pi=2\times\SI{3.8}{kHz}$ in the loading sequence (\hyperref[fig:2]{FIG. 2a}). At the end of a shaking waveform, absorption imaging follows to detect either the atoms centre-of-mass (c.m.) position or the fraction of atoms in the different energy bands ($s,p,d$). The c.m.~position is measured with an \emph{in-situ} image while the atoms are still confined in the optical potentials. The fraction of atoms in each band is determined via the band mapping technique followed by a time-of-flight expansion.

\subsection{\emph{In-situ} detection}
The \emph{in-situ} image is taken right after a shaking waveform is completed, i.e.~in the presence of the lattice, optical trap and homogenous magnetic field. The absorption imaging axis is along $y$ direction and consequently the image captures a two-dimensional (2D) profile of the atoms in the \emph{x-z} plane. The image pixel spacing is \SI{16}{\mu m} and the magnification is 3.8, which gives us a single-pixel resolution of $\SI{4.2}{\mu m}\ \approx 7.9\,a$. To extract the c.m.~and size (defined as the full width at half maximum) of the cloud we use a 2D Gaussian fit. We measure the stability of the c.m.~position in a typical experimental sequence over 100 repetitions and obtain a standard deviation of 0.11 pixels (0.3 pixels peak-to-peak of all repetitions).

\subsection{Band-mapping detection}
The band mapping is implemented with an exponential ramp to switch-off the lattice beam in \SI{500}{\mu s}. Afterwards, the optical trap and also the homogeneous magnetic field are suddenly switched off. A time-of-flight expansion follows during \SI{25}{ms} before absorption imaging. The absorption image is divided into the first, second and third Brillouin zone (BZ), which delineate the regions where atoms belong to the $s$, $p$ or $d$ band, respectively. The BZ size and centre are calibrated with an $s$-band state, where the filling edges are determined by fitting the atomic cloud with an error function. The BZ size is estimated to be 130 pixels wide along the $x$ direction. With this calibration we have an $s$-band insulating state (right after loading into the $4\ E_R$ lattice) consisting of $94.7(4)\%$, $5.7(2)\%$ and $-0.5(2)\%$ of the atoms in the first, second and third BZ, respectively. 

\subsection{Lattice shaking}
Shaking the lattice potential $V(x)=V_0\cos^2(\pi x/a)$ generates the time-dependent ($\tau$) lattice potential $V[x-x_0(\tau)]$, where $x_0(\tau)$ is the shaking waveform.
The optical lattice is formed as one arm of a Michelson interferometer, allowing us to calibrate the amplitude and phase of a cosine waveform for driving frequencies in the range of 1-\SI{100}{kHz} and amplitudes up to several \SI{100}{nm}. We use an arbitrary waveform generator (AWG) to program the two-frequency shaking waveform $x_0(\tau)=\frac{h}{2\pi am}\left(\frac{K_1}{\omega}\cos(\omega\tau)+\frac{K_2}{2\omega}\cos(2\omega\tau+\varphi)\right)$. The AWG allows us to perform time-modulated frequency $\omega(\tau)$, amplitudes $K_{1,2}(\tau)$ and phase $\varphi(\tau)$, which are needed to implement the loading and/or pumping schemes.

\subsection{Lattice depth calibration}
We make use of the lattice shaking to carry out the lattice depth calibration, which consists in \emph{s-p} bandgap spectroscopy. To do this, we load the atoms in the $s$-band and shake the lattice with a simple cosine waveform $x_0(\tau)=\frac{h}{2\pi am}K_1\cos(\omega\tau)$, at constant amplitude $K_1=0.05$. We measure the fraction of atoms excited to the $p$-band for a range of frequencies that span the expected transition resonance at quasi-momentum $q=\pm\pi/2a$. This is repeated for 20, 15, 12.5, 10 and $8\ E_R$ lattice depths and shaking times of 0.6, 0.6, 1, 1 and \SI{4}{ms}, respectively (in addition to switch-on/off times each of \SI{100}{\mu s}). After this procedure, we determine the laser power to produce an optical lattice of 6 or $4\ E_R$ using a linear extrapolation. 

\subsection{Alignment and magnetic gradient calibration}
Detecting a c.m.~shift of $2$ lattice sites after $2$ pump cycles is experimentally demanding, given a single-pixel resolution of $7.9\,a$ per pixel. In order to reduce systematic errors on the c.m.~position measurement, we perform two different optimisations. First, we align the lattice beam to the optical trap  to avoid displacements induced by dipole oscillations. Second, we calibrate out residual magnetic field gradients that can cause an acceleration on the atoms due to the Zeeman effect. For this, we load and shake the atoms with a waveform of constant amplitudes $K_1=0.8, K_2=0.33$ and phase $\varphi=0$, which should not cause any c.m.~shift (symmetric band structure). We repeat this sequence for different magnetic gradients along $x$ produced by an extra coil to compensate a small residual gradient from the pair of coils that generate the homogeneous field of \SI{10}{G}. For each configuration we shake the atoms during 10-\SI{20}{ms} and measure linear c.m.~shifts. The optimal configuration gives a residual slope of the c.m.~shift is $5(5)\times10^{-3}$ pixels/ms.

\subsection{Shaking relative phase $\varphi$ calibration}
We corroborate the shaking phase $\varphi$ calibration inherited from the lattice shaking performed with single-frequency cosine waveforms. The same experimental sequence as for the magnetic gradient calibration is used, i.e.~measure the c.m.~position after \SI{10}{ms} of shaking with constant amplitudes $K_1$ and $K_2$, but scan over the relative phase $\varphi=-\pi, \cdots, +\pi$ in steps of $\pi/6$ with 6 repetitions each. We do this for three different configurations with $K_2=0.07,0.2,0.33$ and $K_1=0.8$. For each data set we fit a sine curve $f(\varphi)=y_0+A\sin(\varphi-\varphi_c)$, with free parameters $y_0,A,\varphi_c$. The standard deviation of the offset $y_0$ out of the three configurations is 0.03 pixels, i.e. the dependence on the $K_2$ amplitude is negligible. The centre relative phase $\varphi_c$ fitted value is $0.01(2)\ \pi$, which agrees with the lattice shaking calibration.

\subsection{Floquet-Bloch bands and choice of shaking parameters}
The computation of numerically exact Floquet-Bloch bands guides us in our experimental work.
The procedure has been described in our previous works ~\cite{viebahn_suppressing_2021,sandholzer_floquet_2022} and in the review by \citet{holthaus_floquet_2016}.
While the relevant physical phenomena are captured by an effective two-band model, the knowledge of the full Floquet band structure is important for practical purposes.
Namely, we can determine spurious $d$-band admixtures, calculate the \emph{s-p} bandgap size for different shaking parameters ($K_{1,2},\varphi$), and identify the critical amplitude $K_2$ for closing the gap when $\varphi=\pm\pi/2$.
A compromise between these specifications led us to choose a lattice depth of $4\ E_R$, $\omega/2\pi=\SI{6.5}{kHz}$ and $K_1=0.8$, which sets the critical points at $K_2=0.22$ and $\varphi=\pm\pi/2$.

In principle, the Floquet pumping scheme is independent of the lattice depth.
In practice, the lattice depth is a trade-off between large band gaps (faster dynamics) and spurious $d$-band admixture.
Shallow lattice depths lead to stronger higher-band coupling~\cite{sandholzer_floquet_2022}.
On the one hand, a large coupling to the $p$-band is advantageous, leading to fast experimental timescales and large topological gaps.
On the other hand, coupling to the $d$-band leads to atom loss and represents an experimental limitation.
Therefore, we choose a lattice depth of $4\,E_R$, which is a compromise between the two limits.
We have measured the fraction of atoms in the $d$-band for all measurements presented in the main text, see \hyperref[sec:suppIII]{III. Additional measurements}.

\subsection{Expansion dynamics of the atomic cloud}
The relevant Floquet physics starts when the shaking waveform is initiated.
Prior to shaking, we ramp down the confining potentials in the lattice direction and reduce the initial depth of $6\,E_R$ to $4\,E_R$, as described previously.
This procedure induces an expansion of the atomic cloud in real space.
Consequently, the initial distribution of quasi-momentum states can change in time, which is evident from the band mapping images in the Floquet state preparation \emph{In-Out} sequence (\hyperref[fig:2]{FIG. 2i}). 
Here, the expansion time, defined as the time elapsed from the moment that the atoms are released into the $4\,E_R$ lattice until the absorption image is taken, is around \SI{14}{ms} when the frequency ramp time is set to \SI{5}{ms}.
In the undriven lattice, during the first \SI{20}{ms} of expansion time we perform a linear fit to estimate the expansion rate and measure 2.6(1) lattice sites per ms ($a/\text{ms}$).
For example, the cloud size is measured to be $140(3)\,a$ and $165(4)\,a$ after expansion times of \SI{10}{ms} and \SI{20}{ms}, respectively.
In contrast, the Floquet state preparation sequence (\hyperref[fig:2]{FIG. 2c}, with shaking relative phase $\varphi=0$) leads to an initial contraction of the atomic cloud, which is then followed by an expansion.
Specifically, the cloud size is $18(3)\,a$ smaller than when measured in the undriven lattice after an expansion time of \SI{10}{ms}.
After the initial contraction, the cloud expands faster than in the undriven lattice, initially at a rate of $2.8(4)\,a/\text{ms}$.
While the precise mechanisms for the contraction are not fully understood, it represents a plausible cause of quasi-momentum redistribution measured in band-mapping (\hyperref[fig:2]{FIG. 2i}).

\section{\large II. Numerical simulations}
\label{sec:suppII}
\subsection{Two-band model}
Numerical simulations presented in the main text are based on a Floquet two-band model of $s,p$ bands. Our model is formulated on similar lines as in references \cite{kang_creutz_2020,sandholzer_floquet_2022} and here we derive a Bloch Hamiltonian starting from the real-space description. We start from the time-dependent Hamiltonian $H_{\text{lab}}(\tau)=\frac{p^2}{2m}+V\left(x-x_0(\tau)\right)$ in the frame where the lattice position is driven by the two-frequency waveform $x_0(\tau)=\frac{h}{2\pi am}\left(\frac{K_1}{\omega}\cos(\omega\tau)+\frac{K_2}{2\omega}\cos(2\omega\tau+\varphi)\right)$. The kinetic energy is $p^2/2m$ and the potential energy is $V(x)=V_0\cos^2(\pi x/a)$. In the comoving frame of the shaken lattice, the Hamiltonian transforms into
\begin{equation}
	H(\tau)=\frac{p^2}{2m}+V(x)-F(\tau)x,
	\end{equation}
where the inertial force is given by $F(\tau)=-m\ddot x_0(\tau)$. In the tight-binding picture, we introduce the second quantised annihilation operator $c_{j\alpha}$ at a site $j$ and band $\alpha=s$ or $p$. The kinetic plus potential energy is given by
\begin{equation*}
	\sum_{j\alpha}\left[ \epsilon_{\alpha}c_{j\alpha}^{\dagger}c_{j\alpha} -\sum_k \left(t_{\alpha}^k c_{j\alpha}^{\dagger}c_{j+k\alpha}+\text{h.c.}\right)\right],
	\end{equation*}
with on-site energy $\epsilon_\alpha$ and $t_{\alpha}^k$ the tunnelling to the $k$-th neighbor in the band $\alpha$.
The normalised position operator $\frac{x}{a}$ is given by
\begin{equation*}
	\sum_{j\alpha}\left\{j c_{j\alpha}^{\dagger}c_{j\alpha} + \sum_{\beta\neq\alpha}\left[\eta_{\alpha\beta}^0c_{j\alpha}^{\dagger}c_{j\beta}+\left(\eta_{\alpha\beta}^1c_{j\alpha}^{\dagger}c_{j+1\beta}+\text{h.c.}\right)\right]\right\},
	\end{equation*}
where the terms $\eta_{\alpha\beta}^0,\eta_{\alpha\beta}^1$ couple different bands $\alpha\neq\beta$ on-site and between nearest neighbors, respectively. Numerical values of the different constants are shown in \hyperref[tab:table-name]{TABLE I}.
\begin{table}
{\renewcommand{\arraystretch}{2}
	\begin{tabular}{ llll }
		\hline
		$\epsilon_s=1.706$ & $t_s^1=8.572\text{e-2}$ & $t_s^2=-5.942\text{e-3}$ & $t_s^3=5.963\text{e-4}$\\ 
		\hline
		$\epsilon_p=4.668$ & $t_p^1=-0.462$ & $t_p^2=-0.081$ & $t_p^3=-0.035$ \\
		\hline
		$\eta_{sp}^0=0.184$ & $\eta_{sp}^1=-0.059$ & &\\ 
		\hline
	\end{tabular}
	}
	\caption{\label{tab:table-name}Relevant parameters of the effective two-band model for a lattice depth of $4\,E_R$. The on-site energies $\epsilon_{\alpha}$ and tunnellings $t_{\alpha}^k$  are in units of $E_R$. The couplings $\eta_{sp}^{0,1}$ are dimensionless constants.}
\end{table}
The on-site term $\mathcal{X}=\sum_{j\alpha}j c_{j\alpha}^{\dagger}c_{j\alpha}$ is rotated away into a tunnelling Peierls phase $A(\tau)=-\frac{a}{\hbar}\int_0^{\tau}F(t')dt'$ via the unitary rotation $R(\tau)=\exp\left(iA(\tau)\mathcal{X}\right)$, which transforms the inter-site terms as
\begin{align*} 
	t_{\alpha}^k c_{j\alpha}^{\dagger}c_{j+k\alpha} &\rightarrow  t_{\alpha}^k e^{-ikA(\tau)} c_{j\alpha}^{\dagger}c_{j+k\alpha} \\ 
	\eta_{\alpha\beta}^1c_{j\alpha}^{\dagger}c_{j+1\beta} &\rightarrow \eta_{\alpha\beta}^1 e^{-iA(\tau)}c_{j\alpha}^{\dagger}c_{j+1\beta}
\end{align*}

\subsection{On-site and neighbouring-site coupling}

Different physical mechanisms are responsible for \emph{s-p} coupling via one- and two-photon processes. In the former case, on-site orbitals are coupled, via the matrix element $\eta_{sp}^0$, in a resonant manner with a single $2\hbar\omega$ photon from the driving force $F(\tau)$. The two-photon process, instead, couples nearest-neighbour orbitals via two $\hbar\omega$ photons.
Here, one photon comes from the force term $-a\eta_{sp}^0F(\tau)$, whereas the other $\hbar\omega$ photon comes from the Peierls phase in the tunnelling $t^1e^{-iA(\tau)}$.
Beyond these lowest-order processes, there are additional couplings between $s$ and $p$ orbitals~\cite{sandholzer_floquet_2022}.
Most importantly, the inter-band nearest-neighbour tunnelling enabled by the matrix element $\eta_{sp}^1$ induces both one- and two-photon processes.
Notably, the $\eta_{sp}^1$ term shifts the critical $K_2$ for the individual gap closing (\hyperref[fig:1]{FIG. 1c}) by approximately $50\%$ compared the lowest-order couplings.

\subsection{Quasi-momentum representation}
In the quasi-momentum spinor $\left(c_{qs},c_{qp}\right)^T$ basis, with $q$ ranging from $-\pi/a$ to $+\pi/a$, the rotated Hamiltonian can be written in terms of the Pauli matrices $\sigma_{x,y,z}$ as the following Bloch Hamiltonian
\begin{equation}
	\begin{split}
		\mathcal{H}(q,\tau)&=\left(\overline{\epsilon}-2\sum_kt_{+}^k\cos(kqa-kA)\right)\mathds{1}\\
		&-aF(\tau)\left(\eta_{sp}^0+2\eta_{sp}^1\cos(qa-A)\right)\sigma_x\\
		&\left(\epsilon-2\sum_kt_{-}^k\cos(kqa-kA)\right)\sigma_z
		\end{split}
	\end{equation}
with $2\overline{\epsilon}=\epsilon_p+\epsilon_s$, $2\epsilon=\epsilon_p-\epsilon_s$ and $2t_{\pm}^k=t_p\pm t_s$. In summary, we have constructed a Bloch Hamiltonian for the $s$, $p$ bands coupled by the Floquet drive starting from the real-space description. Importantly, we used two approximations by considering up to third neighbour tunnellings $t_{s,p}^{1,2,3}$ and first neighbour inter-band couplings $\eta_{sp}^{0,1}$.

\subsection{Effective Hamiltonian and time dynamics}
An effective time-independent Hamiltonian is derived in a rotating frame of the drive generated by the unitary rotation $U(\tau)=\exp(-i\omega\tau\sigma_z)$. The rotated Hamiltonian $\mathcal{H}'(q,\tau)$ has the same form as in (S2), but one replaces $\sigma_x$ by $\sigma_x\cos(2\omega\tau)-\sigma_y\sin(2\omega\tau)$ and $\epsilon$ by the detuning $-\Delta=\epsilon-\hbar\omega$. We numerically compute the effective Hamiltonian using the high-frequency expansion
\begin{equation}
	H_{\text{eff}}=\mathcal{H}_0+\sum_{l=1}^4 \frac{\left[\mathcal{H}_l,\mathcal{H}_l^{\dagger}\right]}{2\pi l\omega}
	\end{equation}
where $\mathcal{H}_l=\frac{1}{T}\int_0^T \mathcal{H}'(q,\tau)e^{-i2\pi l\omega\tau}d\tau$ for $l=0,\cdots,4$ are the Fourier components of the rotated Hamiltonian and $T=2\pi/\omega$ is the Floquet period. The time integral is computed with a time step of $d\tau = \SI{1}{\mu s}$. This is done for quasi-momentum states $q\in\left[-\frac{\pi}{a},\frac{\pi}{a}\right]$ with a quasi-momentum step of $dq=\frac{\pi}{100a}$ and we compute the time-independent velocity operator $v_{\text{eff}}=\frac{1}{\hbar}\partial_qH_{\text{eff}}$.

Time dynamics is obtained by numerically evolving an initial state $\vert\psi(\tau=0)\rangle$ under the Schr{\"o}dinger equation $i\hbar\partial_\tau\vert\psi(\tau)\rangle=H_{\text{eff}}\vert\psi(\tau)\rangle$ for $\tau>0$, with a time step $d\tau = \SI{10}{\mu s}$. We evaluate the velocity of an insulating state, i.e.~a homogeneous distribution of quasi-momentum states, as
\begin{equation*}
	v(\tau)=\frac{1}{2\pi/a}\int_{-\pi/a}^{\pi/a}\langle\psi(\tau)\vert v_{\text{eff}}\vert\psi(\tau)\rangle dq,
	\end{equation*}
and the centre-of-mass (c.m.) position as
\begin{equation*}
	C(\tau)=\int_0^{\tau}v(\tau') d\tau'.
	\end{equation*}

\subsection{Zak phase}
The Zak phase map in the $(K_2,\varphi)$-plane (\hyperref[fig:1]{FIG. 1d}) is calculated using the effective Hamiltonian (S3) with a $K_2$ step of $5\times10^{-3}$ and $\varphi$ step of $\frac{\pi}{160}$. The Zak phase is calculated as
\begin{equation*}
	\varphi_{\text{Zak}}=-\text{Im}\ln\langle u_{-\frac{\pi}{a}}\vert u_{-\frac{\pi}{a}+dq}\rangle\cdots\langle u_{\frac{\pi}{a}-dq}\vert u_{\frac{\pi}{a}}\rangle,
	\end{equation*} 
with $\vert u_{q}\rangle$ the ground-state of $H_{\text{eff}}$ for $q\in\left[-\frac{\pi}{a},\frac{\pi}{a}\right]$, which is defined by having $s$- and $p$-band character at the BZ centre and edges, respectively. The critical amplitude $K_2=0.22$ is the same as when evaluated using exact diagonalisation of Floquet-Bloch bands~\cite{holthaus_floquet_2016,sandholzer_floquet_2022}.

\subsection{Wannier state}
The Wannier states presented in \hyperref[fig:1]{FIG. 1e} were constructed in the so-called \emph{twisted parallel-transport gauge}, as described in~\cite{vanderbilt_berry_2018}. For this, one starts with the set of ground-states $\{\vert u_q\rangle\}$ of $H_{\text{eff}}$ and creates a new set of \emph{parallel transported} states $\{\vert \overline{u}_{q_j}\rangle\}$ that differ from the previous states only by a complex phase. The choice of phases is made such that $\langle\overline{u}_{q_j}\vert\overline{u}_{q_{j+1}}\rangle$ are real and positive. The index $j=0,\cdots,N$ spans the loop of quasi-momentum states, with $j=0$ and $N(=256)$ to be $q_{0}=0$ and $q_{N}=2\pi/a$, respectively, with a quasi-momentum step of $dq=\frac{\pi}{Na}$. Importantly, the two states $\vert\overline{u}_{q_N}\rangle$ and $\vert\overline{u}_{q_0}\rangle$ represent the same physical state, but differ by exactly the phase factor $\exp\left(-i\varphi_{\text{Zak}}\right)$. This discontinuity at the end of the loop is smoothed out by constructing the \emph{twisted parallel-transport gauge} by applying phase twists to create the new set of states $\{\vert\tilde{u}_{q_j}\rangle\}$ with $\vert\tilde{u}_{q_j}\rangle=\exp\left(-ij\varphi_{\text{Zak}}/N\right)\vert\overline{u}_{q_j}\rangle$. With this, we construct the Wannier function as $w(x)=\frac{1}{N}\sum_{j=0}^{N-1}e^{iq_j x}\tilde{u}_{q_j}(x)$ and the density $\vert w(x)\vert^2$ is shown in \hyperref[fig:1]{FIG. 1d,e}.

\begin{figure}[t!]
	\hspace*{-0.7cm}\includegraphics[width = 1.0\columnwidth]{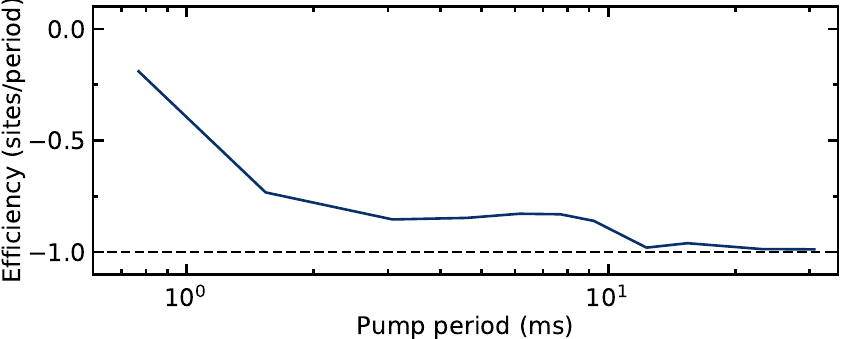}
	\caption{
		Simulated pump efficiency of the Floquet state pumped with topological orbits for pump periods presented in the main text (\hyperref[fig:4]{FIG. 4a}) and longer periods that show quantised efficiency of -1.0 sites per period (horizontal dashed line).
	}
	\label{fig:SMefficiencylong}
\end{figure}

\subsection{Centre-of-mass (c.m.) position dynamics}
The c.m.~position shifts are calculated using the slowly time-modulated effective Hamiltonian $H_{\text{eff}}(\tau)\equiv H_{\text{eff}}(K_2(\tau),\varphi(\tau))$ that describes a pumping orbit. The initial state $\vert\psi(\tau=0)\rangle$ is chosen to be the ground-state of $H_{\text{eff}}(\tau=0)$ (\hyperref[fig:3]{FIG. 3a} and \hyperref[fig:4]{FIG. 4a}) or the $s$-band insulator (\hyperref[fig:3]{FIG. 3b}), which is the eigenstate of the $\sigma_z$ operator with eigen-value $-1$. All pumping orbits are parametrised as
\begin{align*}
	K_2(\tau)&=0.2+0.13\cos(2\pi\tau/\tau_P)\\
	\varphi(\tau)&=\pm\left(\varphi_c-\pi/4\sin(2\pi\tau/\tau_P)\right),
	\end{align*}
with $\varphi_c=+\frac{\pi}{2},0,-\frac{\pi}{2}$ for the red, grey and blue orbits, respectively. The signs $-$ and $+$ in the definition of $\varphi(\tau)$ defines a clock- and counterclockwise orbit, respectively.
The c.m.~position is simulated for up to two pump cycles $\tau\leq2\tau_P$, with periods $\tau_P$ from \SI{0.3}{ms} to \SI{6}{ms} same as presented in the main text.
In addition, we perform a simulation using the blue orbits ($\varphi_c=-\pi/2$), initialised with the ground state of $H_{\text{eff}}(\tau=0)$, and quantised transport is observed only for periods longer than \SI{12}{ms} (FIG. S1).
In addition, we confirm that the pump efficiency is the same when evaluated using the full time-dependent Hamiltonian $\mathcal{H}(q,\tau)$, which was also verified by Kang and Shin~\cite{kang_topological_2020}.
In contrast to simulations based on an effective Hamiltonian, the c.m.~position shift shows micromotion with an amplitude of around 0.15 lattice sites for the same shaking parameters used in the experiment.

\subsection{\emph{In} and \emph{In-Out} measurements of \emph{p}-band population}
Numerical simulations of \emph{In} and \emph{In-Out} measurements presented in the main text (\hyperref[fig:4]{FIG. 4b}) are also calculated using a time-modulated effective Hamiltonian $H_{\text{eff}}(\tau)$. The time dependence is implemented with linear ramps in amplitudes $K_{1,2}(\tau)$ and frequency $\omega(\tau)$ during the loading (and unloading) sequence, and sinusoidal modulations in $K_2(\tau)$ and $\varphi(\tau)$ during the pumping sequence. The initial state $\vert\psi(\tau=0)\rangle$ is always the ground state of $H_{\text{eff}}$ evaluated at $K_2=0.33$, $K_1=0$ and $\omega/2\pi=\SI{3.8}{kHz}$, which is the Hamiltonian prior to start the frequency ramp for the loading sequence. The $p$-band population at time $\tau$ is calculated as $(a/2\pi)\int\vert\langle p\vert \psi(\tau)\rangle\vert^2dq$, where $\vert p\rangle$ is the $p$-band state, which is an eigenstate of the $\sigma_z$ operator with eigen-value $+1$. The simulated amount of $p$-band population is then scaled by the factor (0.947-0.057), and after that it is offset by +0.057. This accounts for our experimental band mapping calibration of the initial $s$-band state as described in \hyperref[sec:suppI]{I. Experimental techniques}. Simulations for the \emph{In-Out} sequence used to optimise the $K_1$ loading time are shown together with experimental data in \hyperref[sec:suppIII]{III. Additional measurements}.

\begin{figure}[t!]
	\hspace*{-0.7cm}\includegraphics[width = 1.0\columnwidth]{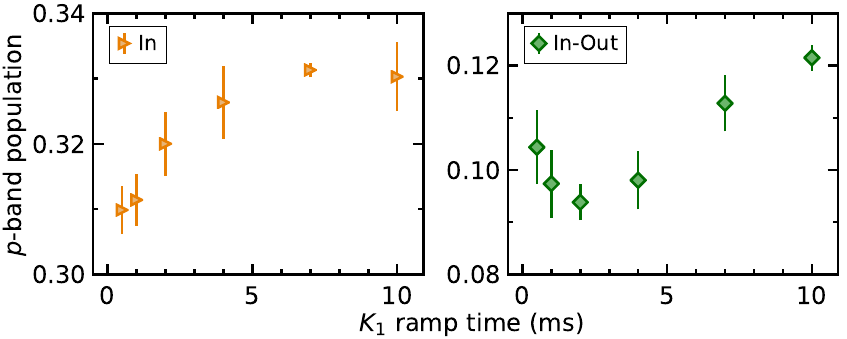}
	\caption{
		$K_1$ ramp optimization, with $p$-band population versus ramp time for \emph{In} (orange triangles) and \emph{In-Out} (green diamonds) sequences. Each data point is the mean and s.e.m. of 3 repetitions.
	}
	\label{fig:SMK1ramp}
\end{figure}

\section{\large III. Additional measurements}
\label{sec:suppIII}
\subsection{\emph{In-Out} measurements for $K_1$ ramp}
The $K_1$ ramp in the loading sequence used to prepare a Floquet state, presented in the main text (\hyperref[fig:2]{FIG. 2a}), was optimised with \emph{In} and \emph{In-Out} sequences (FIG. S2). Here, the \emph{In} sequence consists in a frequency ramp of \SI{5}{ms} followed by the $K_1$ ramp with a variable time and a hold time of \SI{5}{ms}. For the \emph{In-Out} sequence, after the hold time an additional time-reversed $K_1$ ramp and frequency ramp follows. We find a rise of 2\% in $p$-band population saturating at 32\% for the \emph{In} sequence (orange triangles), and a minimum accumulated $p$-band population at \SI{2}{ms} for the \emph{In-Out} sequence (green diamonds). This minimum arises from a combination of two factors. On the one side, more $p$-band atoms accumulate for long ramp times due to decoherence. On the other side, the $p$-band population decreases for sufficiently long, adiabatic ramp times. In addition, we have confirmed that numerical simulations predict an adiabatic time between 1 and \SI{2}{ms}.

\begin{figure}[t!]
	\hspace*{-0.7cm}\includegraphics[width = 1.0\columnwidth]{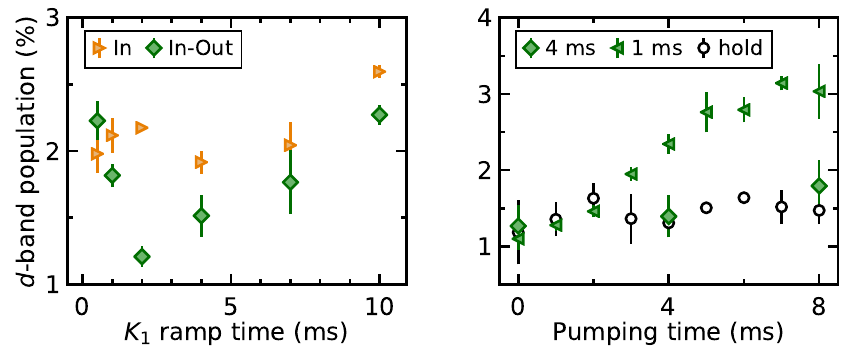}
	\caption{
		Excitations to $d$-band detected in different experiments. \textbf{(left)} $K_1$ ramp optimization for \emph{In} (orange triangles) and \emph{In-Out} (green diamonds) sequences. \textbf{(right)} \emph{In-Out} measurement when pumping with a topological orbit (see \hyperref[fig:4]{FIG. 4b}) with pump periods \SI{4}{ms} (green diamonds), \SI{1}{ms} (green triangles) and hold (black circles). Each data point is the mean and s.e.m. of 3 repetitions.
	}
	\label{fig:SMdband}
\end{figure}

\subsection{Landau-Zener (LZ) transitions in the $2\omega$ ramp}
The $2\omega$ frequency ramp, used for the loading sequence discussed in the main text (\hyperref[fig:2]{FIG. 2f}), can be understood as a LZ transfer. The system consists in a collection of independent two-level systems for each quasi-momentum state $q\in\left[-\frac{\pi}{a},\frac{\pi}{a}\right]$ in which the driving frequency is chirped. The transfer probability from the initial $s$-band to the final Floquet state is given by the LZ formula $P_{\textrm{LZ}}=1-\exp\left(-\pi^2\frac{E_{\textrm{Gap}}^2}{\Delta\nu/\Delta\tau}\right)$. Here, $E_{\textrm{Gap}}$ is the quasi-energy Floquet gap and $\Delta\nu/\Delta\tau$ is the detuning sweep rate. For example, for $q=\pm\pi/2a$ where the Floquet gap is situated after the frequency ramp (\hyperref[fig:2]{FIG. 2d}), we have $E_{\textrm{Gap}}=\SI{0.72}{kHz}$ and a detuning change by $\Delta\nu=2\times\left(6.5-3.8\right)\SI{}{kHz}=\SI{5.4}{kHz}$ during a variable ramp time $\Delta\tau$. For ramp times $\Delta\tau=$\SI{1}{ms}, \SI{5}{ms} and \SI{10}{ms} we deduce $P_{\textrm{LZ}}=$ 0.61, 0.99 and 1.00, respectively. This naive estimate support the experimentally determined adiabatic timescale of \SI{5}{ms} for the frequency ramp (\hyperref[fig:2]{FIG. 2f}). In addition, we have confirmed that numerical simulations predict full transfer from the $s$-band to $p$-band state with a ramp time of \SI{5}{ms}.

\begin{figure}[t!]
	\hspace*{-0.7cm}\includegraphics[width = 1.0\columnwidth]{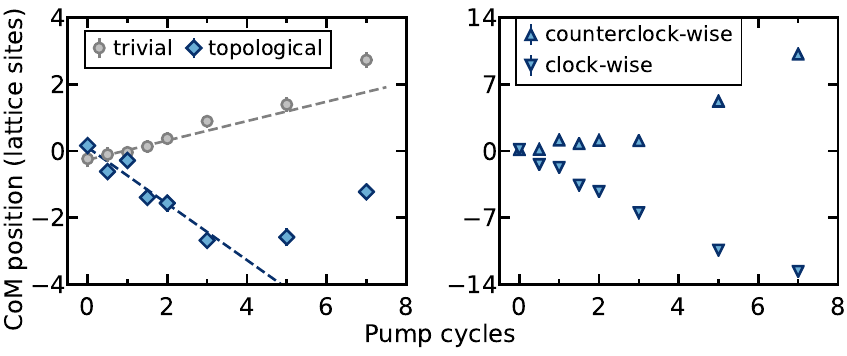}
	\caption{
		Topological pump with \SI{3}{ms} period for long pumping times. \textbf{(left)} Mean centre of mass (c.m.) position when pumping with a topological (blue diamonds) and trivial (grey circles) orbit. This data is used to compute the pump efficiency presented in \hyperref[fig:4]{FIG. 4a}, where dash lines are the linear fits to the c.m.~position versus pump cycles from $0$ to $2$ cycles. \textbf{(right)} c.m.~position shifts for topological clock- (blue inverted triangles) and counterclockwise (blue upright triangles) orbits.
	}
	\label{fig:SMpumplong}
\end{figure}

\subsection{Excitations to the higher $d$-band}
Multi-photon processes can lead to excitations to the higher $d$-band (FIG. S3). In general, $d$-band population is not larger than 3\% for \emph{In} and \emph{In-Out} sequences presented in this work. In the $K_1$ ramp optimization (left panel) the accumulated $d$-band population is minimal at \SI{2}{ms} ramp time for the \emph{In-Out} sequence. This result correlates with the minimum accumulated $p$-band population for that same sequence (see FIG. S2). When pumping with a topological orbit (right panel), we also detected $d$-band population in the \emph{In-Out} sequence (same as in \hyperref[fig:4]{FIG. 4b}). On the one side, pumping with \SI{4}{ms} period (green diamonds) shows accumulated $d$-band population of 1.5\%, constant over the pumping time, which is similar to the hold case (green triangles). On the other side, pumping with \SI{1}{ms} period (black circles) does induce transfers and accumulates up to 3\% $d$-band population. These measurements agree with expected excitations to the $d$-band, estimated via the \emph{s-d} and \emph{p-d} gaps in the numerically exact Floquet-Bloch spectra.

\subsection{Topological pump for long pumping times}
When topological pumping the Floquet state for long times, we find the mean c.m.~position shifts deviating from the near-quantised transport at short times (FIG. S4, left panel). For a pump period of \SI{3}{ms}, the mean c.m.~position shows a non-linear response after 3 pump cycles for the topological (blue diamonds) and trivial (grey circles) orbits. Any factor that breaks the mirror symmetry between the counterclockwise and clockwise orbits can cause this, such as residual magnetic gradients not properly calibrated for long enough times. This would induce c.m.~shifts in the same direction regardless of the pumping orbits for spin-polarised atoms.
This is the reason why we measure the pump efficiency for periods shorter than $6$ ms and pumping times up to 2 cycles in the main text (\hyperref[fig:4]{FIG. 4a}).
We note that for long times the c.m.~position shifts by more than 10 lattice sites and in opposite directions for the individual clockwise and counterclockwise topological orbits (FIG. S4, right panel).
These large displacements can result from an asymmetric filling of the band, or as a filled asymmetric band, or a combination thereof.
In all cases, the asymmetry around $q = 0$ causes significant group velocity effects in addition to the topological transport.

\begin{figure}[t!]
	\hspace*{-0.7cm}\includegraphics[width = 1.0\columnwidth]{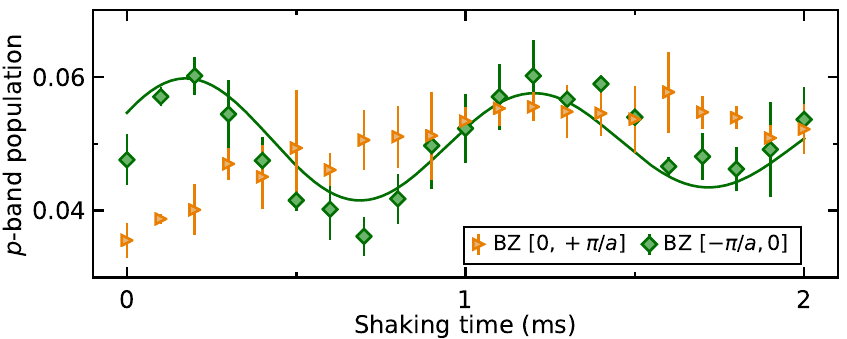}
	\caption{
		Asymmetric \emph{s-p} band Rabi oscillations detected in the $p$-band population of the BZ regions $\left[0,+\pi/a\right]$ (orange triangles) and $\left[-\pi/a,0\right]$ (green diamonds). Green solid line is a damped sinusoidal fit with a fitted period of \SI{1.03}{ms}.
	}
	\label{fig:SMquenchRabi}
\end{figure}

\subsection{Rabi oscillations between \emph{s}- and \emph{p}-band}
A fast switch-on of the shaking leads to \emph{s-p} band Rabi oscillations, which are asymmetric for a two-frequency drive that breaks time-reversal symmetry (FIG. S5). Band mapping images allow quasi-momentum resolved detection of $p$-band population between the BZ regions $\left[0,+\pi/a\right]$ (orange triangles) and $\left[-\pi/a,0\right]$ (green diamonds). In the latter, we measure an oscillation period of \SI{1.03}{ms} (solid green line). This approximately corresponds to the inverse of the Floquet gap $1/\SI{1.19}{kHz}\approx\SI{0.84}{ms}$ at $q=-\pi/2a$ for a shaking waveform with $\omega/2\pi=\SI{6.5}{kHz}$, $K_1=0.8$, $K_2=0.33$ and $\varphi=+\pi/2$ (\hyperref[fig:2]{FIG. 2e}). A complete oscillation is not visible in the region $\left[0,+\pi/a\right]$, where a period of $1/\SI{0.26}{kHz}\approx\SI{3.85}{ms}$ is expected from the inverse of the Floquet gap at $q=+\pi/2a$. These oscillations govern the short-time position dynamics when pumping without loading, which result in zero mean c.m.~position shifts regardless of the pumping orbit (\hyperref[fig:3]{FIG. 3b}).

\subsection{Floquet state loading with a less insulating state}
Loading a Floquet state in a lattice with a less insulating state yields a smaller \emph{s-p} band hybridization (FIG. S6 left panel). For the \emph{In} sequence of the frequency ramp we measure the $p$-band population versus ramp time. We compare the cases when spin-polarised fermions are prepared directly in a $4\ E_R$ lattice (brown diamonds) or first in a $6\ E_R$ lattice which is ramped down to $4\ E_R$ (orange triangles). We find a difference of around $10\%$ in $p$-band population for ramp times longer than \SI{5}{ms}. This shows that using a lattice loading sequence that produces a near-insulating state renders a larger \emph{s-p} hybridisation during the frequency ramp.

\begin{figure}[t!]
	\hspace*{-0.7cm}\includegraphics[width = 1.0\columnwidth]{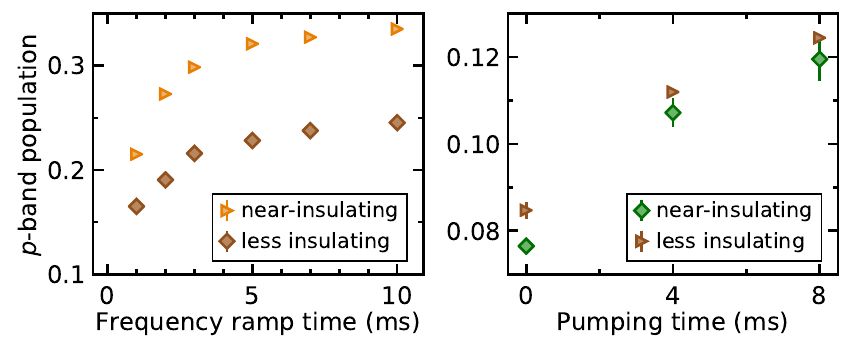}
	\caption{
		\textbf{(left)} Floquet state $p$-band population versus ramp time measured with the \emph{In} sequence of the frequency ramp, for a lattice with near-insulating (orange triangles, same as in \hyperref[fig:2]{FIG. 2f}) and less insulating states (brown diamonds). \textbf{(right)} \emph{In-Out} measurement of $p$-band population versus pumping time for a topological orbit with pump periods \SI{4}{ms} in a lattice with near-insulating (green diamonds, same as in \hyperref[fig:4]{FIG. 4b}) and less insulating states (brown triangles). Each data point is the mean and s.e.m. of 3 repetitions.
	}
	\label{fig:SMpbandOLX44}
\end{figure}

\begin{figure}[t!]
	\hspace*{-0.7cm}\includegraphics[width = 1.0\columnwidth]{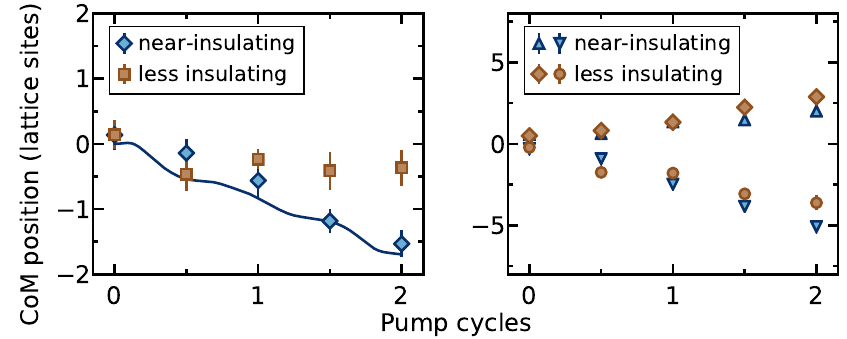}
	\caption{
		Pumping with topological orbits for a pump period of \SI{4}{ms}. \textbf{(left)} Mean centre-of-mass (c.m.) position shifts  in a lattice with near-insulating (blue diamonds) and less insulating states (brown squares). The data for the near-insulating state and the numerical simulation (blue solid line) are reproduced from \hyperref[fig:3]{FIG. 3a}. \textbf{(right)} c.m.~position shifts for individual clockwise and counterclockwise orbits in a lattice with near-insulating (blue upright and inverted triangles, same as in \hyperref[fig:3]{FIG. 3a inset}) and less insulating states (brown diamonds and circles).
	}
	\label{fig:SMpumpOLX44}
\end{figure}

\subsection{Topological pump with a less insulating state}
Topological pumping a Floquet state with a less insulating state results in a smaller pump efficiency (FIG. S7 left panel). We compare the mean c.m.~position shifts (left panel) for the cases when spin-polarised fermions are prepared directly in a $4\ E_R$ lattice (brown squares) or first in a $6\ E_R$ lattice which is ramped down to $4\ E_R$ (blue diamonds). While the latter shows a near-quantised pump in agreement with simulations (solid blue line), a lattice with a less insulating state shows no pumping effect. We shall contrast this result with two additional observables that show no dependence on the filling of the BZ. First, the individual clockwise and counterclockwise topological orbits lead to comparably large c.m.~position shifts (FIG. S7 right panel). Second, the pump adiabaticity measured with an \emph{In-Out} sequence shows similar accumulated $p$-band population versus pumping time (FIG. S6 right panel). The absence of topological pumping in a lattice with a less insulating state, therefore, can be attributed to a smaller $p$-band population of the prepared Floquet state compared to the case when using a near-insulating state (see FIG. S6 left panel).

\clearpage
\end{document}